\documentclass[12pt]{article}
\usepackage{graphicx}
\usepackage{amsmath}
\usepackage{amssymb}
\usepackage{caption2}
\begin{document}
\bibliographystyle{prsty}
\begin{center}
{\large {\bf \sc{   Scalar form-factor of the proton with light-cone QCD sum rules }}} \\[2mm]
Zhi-Gang Wang$^{1}$ \footnote{Corresponding author; E-mail,wangzgyiti@yahoo.com.cn.  }, Shao-Long Wan$^{2}$ and Wei-Min Yang$^{2} $    \\
$^{1}$ Department of Physics, North China Electric Power University, Baoding 071003, P. R. China \\
$^{2}$ Department of Modern Physics, University of Science and Technology of China, Hefei 230026, P. R. China \\
\end{center}

\begin{abstract}
In this article,  we  calculate the
  scalar form-factor of the proton in the framework of the
 light-cone QCD sum rules  approach with the three valence  quark light-cone distribution amplitudes
  up to twist-6, and observe the scalar form-factor $\sigma(t=-Q^2)$ at
  intermediate and large
momentum transfers $Q^2> 2GeV^2$ has significant  contributions from
the end-point (or soft) terms.  The numerical values for the
$\sigma(t=-Q^2)$ are  compatible  with the calculations from the
chiral quark model and lattice QCD at the
 region  $Q^2>2GeV^2$.
 \end{abstract}


\section{Introduction}
The pion-nucleon sigma-term $\Sigma_{\pi N}$ measures the nucleon
mass shift away from the chiral limit and is particularly suited to
test our understanding of the mechanism of the spontaneous and
explicit chiral symmetry breaking  in QCD due to the non-zero $u$,
$d$ quark masses ( For an elegant review of the earlier works, one
can consult Ref.\cite{Reya74} ). The precise knowledge of the values
of the $\Sigma_{\pi N}$ is  of great importance for many
phenomenological applications, for example, the $\Sigma_{\pi N}$
enters the  counting rates in searching for the Higgs boson
\cite{Cheng88}, supersymmetric particles \cite{Bottino99} and dark
matter \cite{Chattopadhyay01,Prezeau03}. However, no experimental
method can be used to  measure the $\Sigma_{ \pi N}$ directly. The
low energy theorem relates the nucleon scalar form-factor
$\sigma(t)$ to the isospin-even $\pi N$ scattering amplitude
$D^{+}(\nu,t)$ at the un-physical Cheng-Dashen point, $\nu=0,
t=2m_{\pi}^2$ \cite{chen71}. The Cheng-Dashen point lies outside the
physical $\pi N$ scattering region,  we have to extrapolate the
experimental $\bar{D}^{+}$ amplitude  to obtain the $\Sigma_{\pi N}$
with the general techniques of the
  dispersion relation and
partial-waves analysis, the bar over $\bar{D}^{+}$ indicates that
the pseudo-vector Born term has been subtracted. Earlier analysis
performed by Koch \cite{Koch82} and Gasser, Leutwyler, Sainio
\cite{Gasser90} gave the canonical value for the $\sigma(2m_\pi^2)$,
 $\sigma(2m_\pi^2)\approx 60\,{\rm MeV}$, however, the recent
analysis of the $\pi N$ scattering data supports the values
$\Sigma_{\pi N} = 79\pm7$ MeV \cite{review2002}. Although there have
been a lot of works on the pion-nucleon sigma-term, for example,
chiral perturbation theory \cite{BKM9396,Borasoy9679,OW99}, lattice
QCD \cite{DLL96,Guesken99,Thomas004,Procura04}, various chiral quark
models \cite{BRS88,DPP89,LGFD01,Schw041}, or Schwinger-Dyson
equation \cite{roberts05}, the value of the sigma term remains  a
puzzle.

In this article, we calculate the scalar  form-factor $\sigma(t)$ of
the proton in the framework of the light-cone sum rules (LCSR)
\cite{LCSR, LCSRreview} which combine  the standard techniques  of
the QCD sum rules with the conventional parton distribution
amplitudes describing  the hard exclusive processes \cite{SVZ79}. In
the LCSR approach, the short-distance  operator product expansion
with   the vacuum condensates of increasing dimensions is replaced
by the light-cone expansion with the distribution amplitudes (which
correspond to the sum of an infinite series of operators with the
same twist) of increasing twists  to parameterize the
non-perturbative  QCD vacuum, while the  contributions from  the
hard re-scattering can be correctly incorporated  as the
$O(\alpha_s)$ corrections \cite{BKM00}. In recent years, there have
been a lot of applications of the LCSR to the mesons, for example,
the form-factors, strong coupling constants and hadronic matrix
elements \cite{LCSRreview}, the applications to the baryons are
cumbersome
 and only the nucleon electromagnetic form-factors
 \cite{BaryonBraun} and the weak decay $\Lambda_b\to
p\ell\nu_\ell$  \cite{Huang04} are studied,  the higher twists
distribution amplitudes   for the baryons were not available until
recently \cite{BFMS}.

The article is arranged as follows:  we derive the light-cone sum
rules for the scalar form-factor $\sigma(t)$ of the proton in
section II; in section III, numerical results and discussion;
section VI is reserved for conclusion.

\section{Light-cone sum rules for the scalar form-factor}
In the following, we write down  the two-point correlation function
$\Pi(P,q)$ in the framework of the LCSR approach,
\begin{eqnarray}
 \Pi(P,q) = i \int d^4 x \, e^{i q \cdot x}
\langle 0| T\left\{\eta(0) J (x)\right\} |P\rangle ,
\end{eqnarray}
 with the scalar current
\begin{eqnarray}
 J(x)=  \bar{u}(x)  u(x) +  \bar{d}(x) d(x) ,
\end{eqnarray}
 and the baryon current
\cite{Che84}
\begin{eqnarray}
 \eta(0) &=& \epsilon^{ijk}
\left[u^i(0) C \!\not\!{z} u^j(0)\right] \, \gamma_5 \!\not\!{z}
d^k(0) \,,
\nonumber\\
 \langle0| \eta(0)  |P\rangle & = & f_{\rm N}\,
(P \cdot z) \!\not\!{z} N(P)\, ,
\end{eqnarray}
 here $z$ is a light-cone vector, $z^2 = 0$, and the  $f_N$ is the
 coupling constant for
  the leading twist distribution amplitude
\cite{earlybaryon}.  At the large Euclidean momenta ${P'}^2=(P-q)^2$
and $q^2 = - Q^2$, the correlation function $\Pi(P,q) $ can be
calculated in perturbation theory. In calculation, we need the
following light-cone expanded quark propagator \cite{BB89},
\begin{eqnarray}
S(x) &=& \frac{i\Gamma(d/2)\not\!x}{2\pi^2(-x^2)^{d/2}} \nonumber
\\
&+& \frac{i\Gamma(d/2-1)}{16\pi^2(-x^2)^{d/2-1}}\int\limits_0^1dv
\Big\{(1-v)\not\!x \sigma_{\mu \nu} G^{\mu\nu}(vx) + v\sigma_{\mu
\nu} G^{\mu\nu}(vx)\!\not\! x \Big\} + ...,
 \end{eqnarray} where $
G_{\mu\nu}= g_sG_{\mu\nu}^a(\lambda^a/2)$ is the gluon field
strength tensor and $d$ is the space-time dimension.  The
contributions proportional to the $G_{\mu\nu}$ can give rise to
four-particle (and five-particle) nucleon distribution amplitudes
with a gluon or quark-antiquark pair in addition to the three
valence quarks, their corrections are usually not expected to play
any significant roles \cite{DFJK} and neglected here
\cite{BaryonBraun,Huang04}. Employ the  light-cone quark propagator
in the correlation function $\Pi(P,q)$, we obtain
\begin{eqnarray}
 \Pi(P,q) &=& i \int d^4 x \, \frac{e^{i q \cdot x}}{2 \pi^2 x^4}
\left\{2(C \!\not\!{z}\!\not\!{x})^{\alpha\beta}
(\gamma_5\!\not\!{z})^\gamma \epsilon^{ijk}\langle 0| T\left\{
u^i_\alpha(0) u^j_\beta(x)d^k_\gamma(0)\right\} |P\rangle\right.
\nonumber \\
&&\left.+(C \!\not\!{z})^{\alpha\beta}
(\gamma_5\!\not\!{z}\!\not\!{x})^\gamma \epsilon^{ijk}\langle 0|
T\left\{ u^i_\alpha(0) u^j_\beta(0)d^k_\gamma(x)\right\}
|P\rangle\right\} \, .
\end{eqnarray}

In the light-cone limit $x^2\to 0$, the remaining three-quark
operator sandwiched between the proton state and the vacuum can be
written in terms of the  nucleon distribution amplitudes
\cite{earlybaryon,Che84,BFMS}. It is obviously that if we only take
into account the three valence quark component of the distribution
amplitudes, the correlation  function
 $i \int d^4 x \, e^{i q \cdot x}
\langle 0| T\left\{\eta(0) \bar{s}(x)s(x) \right\} |P\rangle $ must
be zero, i.e. the strange-component of the  scalar form-factor
$\langle P'|\bar{s}(0)s(0)|P\rangle =0$;  if the strange-component
of the $\pi N$ sigma term manifests  itself, the distribution
amplitudes with additional valence gluons and quark-antiquark pairs
must play significant roles.  The three valence quark components of
the nucleon distribution amplitudes are defined by the matrix
element,
\begin{eqnarray}
&&4\langle0|\epsilon^{ijk}u_\alpha^i(a_1 x)u_\beta^j(a_2
x)u_\gamma^k(a_3
x)|P\rangle=\mathcal{S}_1MC_{\alpha\beta}(\gamma_5N)_{\gamma}+
\mathcal{S}_2M^2C_{\alpha\beta}(\rlap/x\gamma_5N)_{\gamma}\nonumber\\
&&{}+ \mathcal{P}_1M(\gamma_5C)_{\alpha\beta}N_{\gamma}+
\mathcal{P}_2M^2(\gamma_5C)_{\alpha\beta}(\rlap/xN)_{\gamma}+
(\mathcal{V}_1+\frac{x^2M^2}{4}\mathcal{V}_1^M)(\rlap/PC)_{\alpha\beta}(\gamma_5N)_{\gamma}
\nonumber\\&&{}+
\mathcal{V}_2M(\rlap/PC)_{\alpha\beta}(\rlap/x\gamma_5N)_{\gamma}+
\mathcal{V}_3M(\gamma_\mu
C)_{\alpha\beta}(\gamma^\mu\gamma_5N)_{\gamma}+
\mathcal{V}_4M^2(\rlap/xC)_{\alpha\beta}(\gamma_5N)_{\gamma}\nonumber\\&&{}+
\mathcal{V}_5M^2(\gamma_\mu
C)_{\alpha\beta}(i\sigma^{\mu\nu}x_\nu\gamma_5N)_{\gamma} +
\mathcal{V}_6M^3(\rlap/xC)_{\alpha\beta}(\rlap/x\gamma_5N)_{\gamma}\nonumber\\&&{}
+(\mathcal{A}_1+\frac{x^2M^2}{4}\mathcal{A}_1^M)(\rlap/P\gamma_5
C)_{\alpha\beta}N_{\gamma}+
\mathcal{A}_2M(\rlap/P\gamma_5C)_{\alpha\beta}(\rlap/xN)_{\gamma}+
\mathcal{A}_3M(\gamma_\mu\gamma_5 C)_{\alpha\beta}(\gamma^\mu
N)_{\gamma}\nonumber\\&&{}+
\mathcal{A}_4M^2(\rlap/x\gamma_5C)_{\alpha\beta}N_{\gamma}+
\mathcal{A}_5M^2(\gamma_\mu\gamma_5
C)_{\alpha\beta}(i\sigma^{\mu\nu}x_\nu N)_{\gamma}+
\mathcal{A}_6M^3(\rlap/x\gamma_5C)_{\alpha\beta}(\rlap/x
N)_{\gamma}\nonumber\\&&{}+(\mathcal{T}_1+\frac{x^2M^2}{4}\mathcal{T}_1^M)(P^\nu
i\sigma_{\mu\nu}C)_{\alpha\beta}(\gamma^\mu\gamma_5
N)_{\gamma}+\mathcal{T}_2M(x^\mu P^\nu
i\sigma_{\mu\nu}C)_{\alpha\beta}(\gamma_5
N)_{\gamma}\nonumber\\&&{}+
\mathcal{T}_3M(\sigma_{\mu\nu}C)_{\alpha\beta}(\sigma^{\mu\nu}\gamma_5
N)_{\gamma}+
\mathcal{T}_4M(P^\nu\sigma_{\mu\nu}C)_{\alpha\beta}(\sigma^{\mu\rho}x_\rho\gamma_5
N)_{\gamma}\nonumber\\&&{}+ \mathcal{T}_5M^2(x^\nu
i\sigma_{\mu\nu}C)_{\alpha\beta}(\gamma^\mu\gamma_5 N)_{\gamma}+
\mathcal{T}_6M^2(x^\mu P^\nu
i\sigma_{\mu\nu}C)_{\alpha\beta}(\rlap/x\gamma_5
N)_{\gamma}\nonumber\\&&{}+
\mathcal{T}_7M^2(\sigma_{\mu\nu}C)_{\alpha\beta}(\sigma^{\mu\nu}\rlap/x\gamma_5
N)_{\gamma}+
\mathcal{T}_8M^3(x^\nu\sigma_{\mu\nu}C)_{\alpha\beta}(\sigma^{\mu\rho}x_\rho\gamma_5
N)_{\gamma} \, \, .
\end{eqnarray}
The calligraphic distribution amplitudes do not have definite twist
and can be related to the ones with definite twist as
\begin{eqnarray}
&&\mathcal{S}_1=S_1, \hspace{0.8cm}2P\cdot
x\mathcal{S}_2=S_1-S_2,\nonumber\\&& \mathcal{P}_1=P_1,
\hspace{0.8cm}2P\cdot x\mathcal{P}_2=P_1-P_2 \nonumber
\end{eqnarray}
for the scalar and pseudo-scalar distribution amplitudes,
\begin{eqnarray}
&&\mathcal{V}_1=V_1, \hspace{2.4cm}2P\cdot
x\mathcal{V}_2=V_1-V_2-V_3, \nonumber\\&& 2\mathcal{V}_3=V_3,
\hspace{2.2cm} 4P\cdot
x\mathcal{V}_4=-2V_1+V_3+V_4+2V_5,\nonumber\\&& 4P\cdot
x\mathcal{V}_5=V_4-V_3,\hspace{0.5cm} (2P\cdot
x)^2\mathcal{V}_6=-V_1+V_2+V_3+V_4+V_5-V_6 \nonumber
\end{eqnarray}
for the vector distribution amplitudes,
\begin{eqnarray}
&&\mathcal{A}_1=A_1, \hspace{2.4cm}2P\cdot
x\mathcal{A}_2=-A_1+A_2-A_3, \nonumber\\&& 2\mathcal{A}_3=A_3,
\hspace{2.2cm}4P\cdot x\mathcal{A}_4=-2A_1-A_3-A_4+2A_5,
\nonumber\\&& 4P\cdot x\mathcal{A}_5=A_3-A_4,\hspace{0.5cm} (2P\cdot
x)^2\mathcal{A}_6=A_1-A_2+A_3+A_4-A_5+A_6 \nonumber
\end{eqnarray}
for the axial vector distribution amplitudes, and
\begin{eqnarray}
&&\mathcal{T}_1=T_1, \hspace{3.85cm}2P\cdot
x\mathcal{T}_2=T_1+T_2-2T_3, \nonumber\\&&
2\mathcal{T}_3=T_7,\hspace{3.68cm} 2P\cdot
x\mathcal{T}_4=T_1-T_2-2T_7, \nonumber\\&& 2P\cdot
x\mathcal{T}_5=-T_1+T_5+2T_8, \hspace{0.5cm}(2P\cdot
x)^2\mathcal{T}_6=2T_2-2T_3-2T_4+2T_5+2T_7+2T_8, \nonumber\\&& 4P
\cdot x\mathcal{T}_7=T_7-T_8, \hspace{1.90cm}(2P\cdot
x)^2\mathcal{T}_8=-T_1+T_2 +T_5-T_6+2T_7+2T_8 \nonumber
\end{eqnarray}
for the tensor distribution amplitudes. The distribution amplitudes
$F=V_i$, $A_i$, $T_i$, $S_i$, $P_i$ can be represented as
\begin{equation}
F(a_ip\cdot x)=\int \mathcal{D}x e^{-ip\cdot
x\Sigma_ix_ia_i}F(x_i)\; ,
\end{equation}
with
\begin{eqnarray}
\mathcal{D}x=dx_1dx_2dx_3\delta(x_1+x_2+x_3-1). \nonumber
\end{eqnarray}
Those light-cone distribution amplitudes are scale dependent and can
be expanded with the conformal operators, to the next-to-leading
conformal spin accuracy, we obtain \cite{BFMS},
\begin{eqnarray}
V_1(x_i,\mu)&=&120x_1x_2x_3[\phi_3^0(\mu)+\phi_3^+(\mu)(1-3x_3)],\nonumber\\
V_2(x_i,\mu)&=&24x_1x_2[\phi_4^0(\mu)+\phi_3^+(\mu)(1-5x_3)],\nonumber\\
V_3(x_i,\mu)&=&12x_3\{\psi_4^0(\mu)(1-x_3)+\psi_4^-(\mu)[x_1^2+x_2^2-x_3(1-x_3)]
\nonumber\\&&+\psi_4^+(\mu)(1-x_3-10x_1x_2)\},\nonumber\\
V_4(x_i,\mu)&=&3\{\psi_5^0(\mu)(1-x_3)+\psi_5^-(\mu)[2x_1x_2-x_3(1-x_3)]
\nonumber\\&&+\psi_5^+(\mu)[1-x_3-2(x_1^2+x_2^2)]\},\nonumber\\
V_5(x_i,\mu)&=&6x_3[\phi_5^0(\mu)+\phi_5^+(\mu)(1-2x_3)],\nonumber\\
V_6(x_i,\mu)&=&2[\phi_6^0(\mu)+\phi_6^+(\mu)(1-3x_3)],\nonumber\\
A_1(x_i,\mu)&=&120x_1x_2x_3\phi_3^-(\mu)(x_2-x_1),\nonumber\\
A_2(x_i,\mu)&=&24x_1x_2\phi_4^-(\mu)(x_2-x_1),\nonumber\\
A_3(x_i,\mu)&=&12x_3(x_2-x_1)\{(\psi_4^0(\mu)+\psi_4^+(\mu))+\psi_4^-(\mu)(1-2x_3)
\},\nonumber\\
A_4(x_i,\mu)&=&3(x_2-x_1)\{-\psi_5^0(\mu)+\psi_5^-(\mu)x_3
+\psi_5^+(\mu)(1-2x_3)\},\nonumber\\
A_5(x_i,\mu)&=&6x_3(x_2-x_1)\phi_5^-(\mu)\nonumber\\
A_6(x_i,\mu)&=&2(x_2-x_1)\phi_6^-(\mu),\nonumber\\
T_1(x_i,\mu)&=&120x_1x_2x_3[\phi_3^0(\mu)+\frac{1}{2}(\phi_3^--\phi_3^+)(\mu)(1-3x_3)
],\nonumber\\
T_2(x_i,\mu)&=&24x_1x_2[\xi_4^0(\mu)+\xi_4^+(\mu)(1-5x_3)],\nonumber\\
T_3(x_i,\mu)&=&6x_3\{(\xi_4^0+\phi_4^0+\psi_4^0)(\mu)(1-x_3)+
(\xi_4^-+\phi_4^--\psi_4^-)(\mu)[x_1^2+x_2^2-x_3(1-x_3)]
\nonumber\\
&&+(\xi_4^++\phi_4^++\psi_4^+)(\mu)(1-x_3-10x_1x_2)\},\nonumber\\
T_4(x_i,\mu)&=&\frac{3}{2}\{(\xi_5^0+\phi_5^0+\psi_5^0)(\mu)(1-x_3)+
(\xi_5^-+\phi_5^--\psi_5^-)(\mu)[2x_1x_2-x_3(1-x_3)]
\nonumber\\
&&+(\xi_5^++\phi_5^++\psi_5^+)(\mu)(1-x_3-2(x_1^2+x_2^2))\},\nonumber\\
T_5(x_i,\mu)&=&6x_3[\xi_5^0(\mu)+\xi_5^+(\mu)(1-2x_3)],\nonumber\\
T_6(x_i,\mu)&=&2[\phi_6^0(\mu)+\frac{1}{2}(\phi_6^--\phi_6^+)(\mu)(1-3x_3)],
\nonumber \\
T_7(x_i,\mu)&=&6x_3\{(-\xi_4^0+\phi_4^0+\psi_4^0)(\mu)(1-x_3)+
(-\xi_4^-+\phi_4^--\psi_4^-)(\mu)[x_1^2+x_2^2-x_3(1-x_3)]
\nonumber\\
&&+(-\xi_4^++\phi_4^++\psi_4^+)(\mu)(1-x_3-10x_1x_2)\},\nonumber\\
T_8(x_i,\mu)&=&\frac{3}{2}\{(-\xi_5^0+\phi_5^0+\psi_5^0)(\mu)(1-x_3)+
(-\xi_5^-+\phi_5^--\psi_5^-)(\mu)[2x_1x_2-x_3(1-x_3)]
\nonumber\\
&&+(-\xi_5^++\phi_5^++\psi_5^+)(\mu)(1-x_3-2(x_1^2+x_2^2))\},\nonumber\\
S_1(x_i,\mu) &=& 6 x_3 (x_2-x_1) \left[ (\xi_4^0 + \phi_4^0 +
\psi_4^0 + \xi_4^+ + \phi_4^+ + \psi_4^+)(\mu) + (\xi_4^- + \phi_4^-
- \psi_4^-)(\mu)(1-2 x_3) \right] \,,
\nonumber \\
S_2(x_i,\mu) &=& \frac{3}{2} (x_2 -x_1) \left[- \left(\psi_5^0 +
\phi_5^0 + \xi_5^0\right)(\mu) + \left(\xi_5^- + \phi_5^- - \psi_5^0
\right)(\mu) x_3 \right. \nonumber \\
 && \left.+\left(\xi_5^+ + \phi_5^+ + \psi_5^0 \right)(\mu) (1- 2
x_3)\right]\,,
\nonumber \\
P_1(x_i,\mu) &=& 6 x_3 (x_2-x_1) \left[ (\xi_4^0 - \phi_4^0 -
\psi_4^0 + \xi_4^+ - \phi_4^+ - \psi_4^+)(\mu) + (\xi_4^- - \phi_4^-
+ \psi_4^-)(\mu)(1-2 x_3) \right] \, ,
\nonumber \\
P_2(x_i,\mu) &=& \frac32 (x_2 -x_1) \left[\left(\psi_5^0 + \psi_5^0
- \xi_5^0\right)(\mu) + \left(\xi_5^- - \phi_5^- + \psi_5^0
\right)(\mu) x_3 \right. \nonumber\\
&& \left. + \left(\xi_5^+ - \phi_5^+ - \psi_5^0 \right)(\mu) (1- 2
x_3)\right]\, .
\end{eqnarray}
The $V_1$, $A_1$ and $T_1$ are leading twist-3 distribution
amplitudes; the $S_1$, $P_1$, $V_2$, $V_3$, $A_2$, $A_3$, $T_2$,
$T_3$ and $T_7$ are twist-4 distribution amplitudes; the $S_2$,
$P_2$, $V_4$, $V_5$, $A_4$, $A_5$, $T_4$, $T_5$ and $T_8$ are
twist-5 distribution amplitudes; while the twist-6 distribution
amplitudes are the $V_6$, $A_6$ and $T_6$. Those parameters
$\phi_3^0$, $\phi_6^0$, $\phi_4^0$, $\phi_5^0$, $\xi_4^0$,
$\xi_5^0$, $\psi_4^0$, $\psi_5^0$, $\phi_3^-$, $\phi_3^+$,
$\phi_4^-$, $\phi_4^+$, $\psi_4^-$, $\psi_4^+$, $\xi_4^-$,
$\xi_4^+$, $\phi_5^-$, $\phi_5^+$, $\psi_5^-$, $\psi_5^+$,
$\xi_5^-$, $\xi_5^+$, $\phi_6^-$, $\phi_6^+ $ can be expressed in
terms of eight  independent matrix elements of the local operators,
for the details, one can consult Ref.\cite{BFMS}.

Taking into account the three valence quark light-cone distribution
amplitudes up to twist-6 and performing the integration over the $x$
in the coordinate space, finally we  obtain the following results,
\begin{eqnarray}
\Pi(P,q)&=&  \!\not\!{z}P\cdot z N(P)\left\{M\int_0^1dt_2t_2\int_0^{1-t_2}dt_1 \frac{1}{(q-t_2P)^2}\right.\nonumber\\
&&\left.\left\{2\left[S_1+T_7\right](t_1,t_2,1-t_1-t_2)-V_3(t_1,1-t_1-t_2,t_2)\right\}\right.
\nonumber \\
&+&\left.M\int_0^1 d\lambda \int_1^\lambda dt_2 \int_0^{1-t_2}dt_1
\frac{1}{(q-\lambda P)^2}
\right.\nonumber \\
&&\left.\left\{\left[V_1-V_2-V_3\right](t_1,1-t_1-t_2,t_2)
-2\left[T_1-T_3-T_7\right](t_1,t_2,1-t_1-t_2)
\right\}\right. \nonumber \\
 &+&\left.4M\int_0^1 d\lambda \lambda\int_1^\lambda dt_2
\int_0^{1-t_2}dt_1\frac{ (q-\lambda P)\cdot P}{(q-\lambda
P)^4}\left[T_2-T_3+T_7\right](t_1,t_2,1-t_1-t_2)
\right. \nonumber \\
&+&\left.M^3 \int_0^1 d\tau \tau\int_1^\tau d\lambda \int_1^\lambda
dt_2 \int_0^{1-t_2}dt_1 \frac{1}{(q-\tau
P)^4}\right.\nonumber\\
&&\left.\left[-V_1+V_2+V_3+V_4+V_5-V_6\right]
(t_1,1-t_1-t_2,t_2)\right\}
\nonumber\\
&+&\cdots \, .
\end{eqnarray}

According to the basic assumption of current-hadron duality in the
QCD sum rules approach \cite{SVZ79}, we insert  a complete series of
intermediate states satisfying the unitarity   principle with the
same quantum numbers as the current operator $\eta(0)$
 into the correlation function in
Eq.(1)  to obtain the hadronic representation. After isolating the
pole terms of the lowest proton  state, we obtain the following
result,
\begin{eqnarray}
\Pi(P,q)&=&\frac{\!\not\!{z}P'\cdot z f_N N(P-q)\langle
N(P-q)|\bar{u}(0)u(0)+\bar{d}(0)d(0)|N(P)\rangle}{M^2-(q-P)^2}+\cdots
\nonumber \\
&=&\frac{\!\not\!{z}P'\cdot z f_N \left\{ \!\not\!{P}-\!\not\!{q}+M
\right\}\sigma(t)N(P)}{ \hat{m}\left[M^2-(q-P)^2\right]}+\cdots \, ,
\end{eqnarray}
here $\hat{m}=\frac{m_u+m_d}{2}$. The structure $\!\not\!{z}$ has an
odd number of $\gamma$-matrix and conserves chirality, the
structures $\!\not\!{z} \!\not\!{P}$ , $\!\not\!{z} \!\not\!{q}$
have even number of $\gamma$-matrixes and violate chirality. In the
original QCD sum rules analysis of the nucleon magnetic
 moments \cite{Ioffe84}, the interval of dimensions (of the condensates) for the odd
structure is larger than the interval of dimensions for the even
structures, one may expect a better accuracy of the results obtained
from the sum rules with  the odd structure. In this article, we
choose the structure $\!\not\!{z}$ for analysis.

The Borel transformation and the continuum states subtraction can be
performed by using the following substitution rules,
\begin{eqnarray}
\int dx \frac{\rho(x)}{(q-xP)^2}&=&-\int_0^1 \frac{dx}{x}
\frac{\rho(x)}{s-{P'}^2}\Rightarrow -\int_{x_0}^1 \frac{dx}{x}
\rho(x)e^{-\frac{s}{M_B^2}} , \nonumber \\
\int dx \frac{\rho(x)}{(q-xP)^4}&=&\int_0^1 \frac{dx}{x^2}
\frac{\rho(x)}{(s-{P'}^2)^2}\Rightarrow  \frac{1}{M_B^2}\int_{x_0}^1
\frac{dx}{x^2}
\rho(x)e^{-\frac{s}{M_B^2}}+\frac{\rho(x_0)e^{-\frac{s_0}{M_B^2}}}{Q^2+x_0^2
M^2} , \nonumber\\
 s&=&(1-x)M^2+\frac{(1-x)}{x}Q^2, \nonumber\\
x_0&=&\frac{\sqrt{(Q^2+s_0-M^2)^2+4M^2Q^2}-(Q^2+s_0-M^2)}{2M^2}.
\end{eqnarray}
Finally we obtain the sum rule for the scalar form-factor
$\sigma(t=-Q^2)$,
\begin{eqnarray}
&&\sigma(t)f_N e^{-\frac{M^2}{M_B^2}}\nonumber\\
&=& -\hat{m} \int_{x_0}^1dt_2\int_0^{1-t_2}dt_1 \exp \left\{-\frac{t_2(1-t_2)M^2+(1-t_2)Q^2}{t_2M_B^2}\right\}\nonumber\\
&&\left\{2\left[S_1+T_7\right](t_1,t_2,1-t_1-t_2)-V_3(t_1,1-t_1-t_2,t_2)\right\}
\nonumber \\
&-&\hat{m} \int_{x_0}^1 \frac{d\lambda}{\lambda} \int_1^\lambda dt_2
\int_0^{1-t_2}dt_1 \exp
\left\{-\frac{\lambda(1-\lambda)M^2+(1-\lambda)Q^2}{\lambda M_B^2}
\right\}
\nonumber \\
&&\left\{\left[V_1-V_2-V_3\right](t_1,1-t_1-t_2,t_2)
-2\left[T_1+T_2-2T_3\right](t_1,t_2,1-t_1-t_2)\right\} \nonumber \\
 &-&2\hat{m} \int_{x_0}^1 \frac{d\lambda}{\lambda^2} \int_1^\lambda dt_2
\int_0^{1-t_2}dt_1
\frac{Q^2+\lambda^2 M^2}{M_B^2}\nonumber \\
&& \exp \left\{-\frac{\lambda(1-\lambda)M^2+(1-\lambda)Q^2}{\lambda
M_B^2} \right\}\left[T_2-T_3+T_7\right](t_1,t_2,1-t_1-t_2)\nonumber \\
 &-&2 \hat{m}  \int_1^{x_0} dt_2
\int_0^{1-t_2}dt_1
 \exp \left\{-\frac{s_0}{M_B^2} \right\}
\left[ T_2-T_3+T_7\right](t_1,t_2,1-t_1-t_2)  \nonumber \\
&+&\frac{\hat{m} M^2}{M_B^2} \int_{x_0}^1 \frac{d\tau}{\tau}
\int_1^\tau d\lambda \int_1^\lambda dt_2 \int_0^{1-t_2}dt_1 \exp
\left\{-\frac{\tau(1-\tau)M^2+(1-\tau)Q^2}{\tau
M_B^2}\right\}\nonumber \\
&&\left[-V_1+V_2+V_3+V_4+V_5-V_6\right](t_1,1-t_1-t_2,t_2)\nonumber\\
&+&\frac{x_0\hat{m} M^2}{M^2+x_0^2M^2}  \int_1^{x_0} d\lambda
\int_1^\lambda
dt_2 \int_0^{1-t_2}dt_1 \exp \left\{-\frac{s_0}{M_B^2}\right\}\nonumber \\
&&\left[-V_1+V_2+V_3+V_4+V_5-V_6\right](t_1,1-t_1-t_2,t_2) .
\end{eqnarray}

\section{Numerical results and discussions}

The input parameters have to be specified  before the numerical
analysis. We choose the suitable values for the Borel parameter
$M_B$, $M_B^2=(1.5-2.5)GeV^2$. In this range, the Borel parameter
$M_B$ is small enough  to warrant the higher mass resonances and
 continuum states are sufficiently suppressed, on the other hand,  it is
  large enough to warrant the convergence of the
light-cone expansion with increasing twists in the perturbative  QCD
calculations \cite{Ioffe84,Ioffe81}. The numerical results show that
in this range, the scalar form-factor $\sigma(t=-Q^2)$ is almost
independent on the Borel parameter $M_B$, which we can see from the
Fig.1 for $Q^2=3GeV^2$, $4GeV^2$ and $5GeV^2$.
\begin{figure}
 \centering
 \includegraphics[totalheight=7cm]{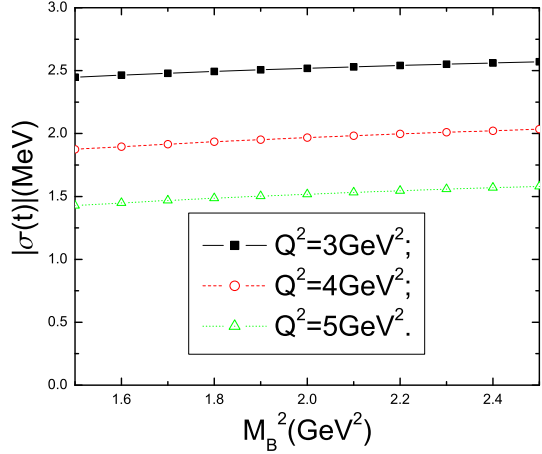}
 \caption{The   $\sigma(t)$ with the Borel parameter $M_B^2$ for $s_0=2.25GeV^2$. }
\end{figure}
In this article, we take the middle point for the Borel parameter,
$M_B^2=2.0GeV^2$ in numerical  analysis, such a specialization will
not lead to  much uncertainties on the final results and impair the
predictive ability.

There are two independent interpolating currents with spin
$\frac{1}{2}$ and isospin $\frac{1}{2}$, both are expected to excite
the ground state proton from the vacuum, the general form of the
proton current can be written as \cite{Baryon}
\begin{eqnarray}
J(x,t) &=&  \epsilon^{abc} \left\{\left[ u_a^T(x) C \gamma_5 d_b(x)
\right]  u_c(x)
 + t \left[ u_a^T(x) C d_b(x)
\right] \gamma_5 u_c(x) \right \},
\end{eqnarray}
in the limit $t=-1$, we recover the Ioffe current. The Monte-Carlo
calculations for the two-point vacuum correlation function indicate
that the optimal  mixing  be $t=-1.2$, the threshold parameter be
$\sqrt{s_0}=(1.53\pm0.41 )GeV$ and the mass of the proton be
$M=(1.17\pm 0.26 ) GeV$ \cite{Leinweber97}, furthermore, the
Monte-Carlo method has been successfully  applied  in studying  the
axial coupling constant and the magnetic moments of the decuplet
baryons \cite{Lee97}. Here we take the Ioffe-type current $\eta(x)$
in Eq.(3) to keep in consistent with the sum rules used in
determining the parameters in the light-cone distribution
amplitudes, $\phi_3^0$, $\phi_6^0$, $\phi_4^0$, $\phi_5^0$,
$\xi_4^0$, $\xi_5^0$, $\psi_4^0$, $\psi_5^0$, $\phi_3^-$,
$\phi_3^+$, $\phi_4^-$, $\phi_4^+$, $\psi_4^-$, $\psi_4^+$,
$\xi_4^-$, $\xi_4^+$, $\phi_5^-$, $\phi_5^+$, $\psi_5^-$,
$\psi_5^+$, $\xi_5^-$, $\xi_5^+$, $\phi_6^-$, $\phi_6^+ $;
 furthermore, we take the physical mass for the proton rather than
determine it from the corresponding sum rules, this obviously leads
to some deviations from the Monte-Carlo adjusted threshold parameter
$s_0$.
\begin{figure}
 \centering
 \includegraphics[totalheight=12cm]{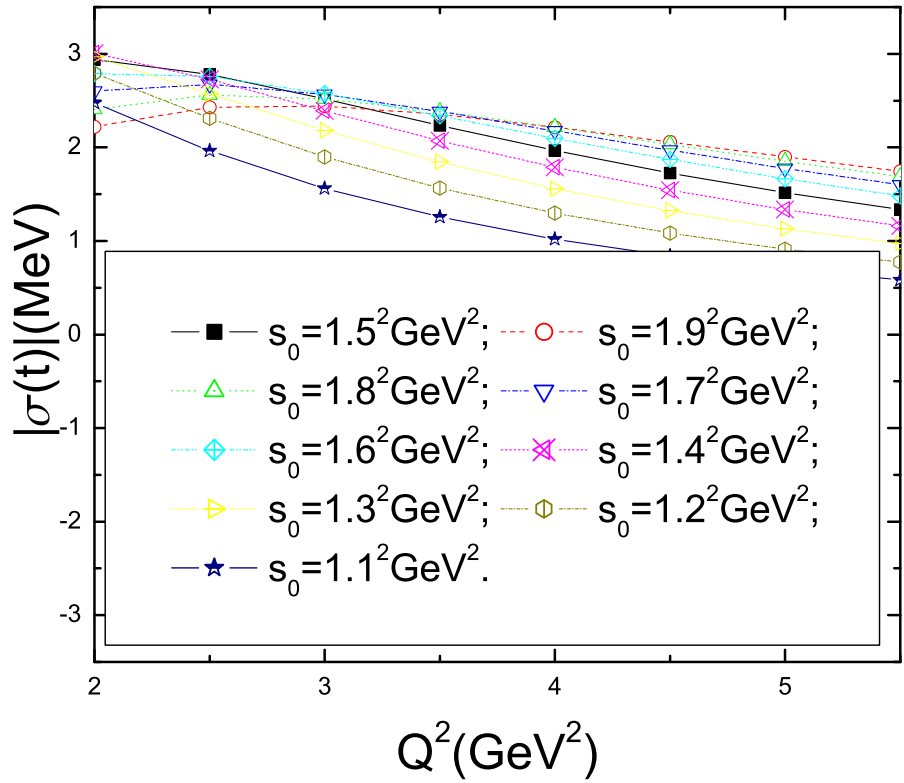}
 \caption{The  $\sigma(t)$ with  the central values of the input parameters.}
\end{figure}
In the Fig.2,  we plot the dependence on the threshold parameter
$s_0$ for the sum rules in a large range,  $\sqrt{s_0}=(1.5\pm
0.4)GeV$. From the figure, we can see that in the region
$Q^2>3GeV^2$, the scalar form-factor is insensitive to threshold
parameter $s_0$, while in the region  $Q^2<3GeV^2$, the curves for
the form-factor $\sigma(t=-Q^2)$ with $\sqrt{s_0}<1.5GeV$ are quite
different from those ones with $\sqrt{s_0}>1.5GeV$, which may be due
to the contributions from the high resonances and continuum states.
For simplicity, we choose the standard value  for the threshold
parameter $s_0$, $s_0=2.25GeV^2$ to subtract the contributions from
the higher resonances and
 continuum states, i.e. we restrict the range of integral  to the energy region
 below the Roper resonance ($N(1440)$);  on the other hand, it is large enough to
include all the contributions from the proton. The current masses
for the $u$ and $d$ quarks are, $m_u=(1.5-4)MeV$ and $m_d=(4-8)MeV$
from the Particle Data Group in 2004 \cite{PDG}, we choose
$\hat{m}=\frac{m_u+m_d}{2}=5MeV$. For $Q^2=(2-5)GeV^2$, $x\geq
x_0=0.5-0.7$, and the average value $\langle x\rangle=0.75-0.85$,
with the intermediate and large space-like momentum $Q^2$, the
end-point contributions (or the Feynman mechanism) are dominant
\footnote{Our work on the axial form-factor of the nucleons with the
LCSR also leads to the same observation, and will be presented
elsewhere. }, it is consistent with the growing consensus that the
onset of the perturbative QCD region in exclusive processes is
postponed to very large energy scales.

 The parameters in the light-cone distribution amplitudes  $ \phi_3^0 $,
  $\phi_6^0$, $\phi_4^0$, $\phi_5^0$,  $\xi_4^0$, $\xi_5^0$, $\psi_4^0$,
$\psi_5^0 $,  $ \phi_3^-$, $\phi_3^+$, $\phi_4^-$, $\phi_4^+$,
$\psi_4^-$, $\psi_4^+$, $\xi_4^-$, $\xi_4^+$, $\phi_5^-$,
$\phi_5^+$, $\psi_5^-$, $\psi_5^+$, $\xi_5^-$ ,$ \xi_5^+$,
$\phi_6^-$, $\phi_6^+ $ are scale dependent and can be calculated
with the corresponding QCD sum rules, the approximated central
values are presented in the Table 1.
\begin{table}
\begin{center}
\begin{tabular}{|l||l|l|l||l|l|l||l|l|l|l}
\hline
         &  $\phi_i^0$  &  $\phi_i^-$ &  $\phi_i^+$
         &  $\psi_i^0$  &  $\psi_i^-$ &  $\psi_i^+$
         &  $\xi_i^0$   &  $\xi_i^-$  &  $\xi_i^+$
\\ \hline
twist-3: $i = 3$
         & $\phantom{-}0.53$         &  $\phantom{-}2.11$       &   0.57
         &              &             &
         &              &             &
\\ \hline
twist-4:  $ i = 4$
         & $-1.08$        &  $\phantom{-} 3.22 $     &   2.12
         & 1.61         &   $-6.13$     &   0.99
         & 0.85         &   $\phantom{-} 2.79$     &  0.56
\\ \hline
twist-5: $i = 5$
         & $-1.08$        &  $-2.01$      &   1.42
         & 1.61         &  $-0.98$      &   -0.99
         & 0.85         &  $-0.95$      &   0.46
\\ \hline
twist-6: $i = 6$
         & $\phantom{-}0.53$         &  $\phantom{-} 3.09$      &   -0.25
         &              &             &
         &              &             &
\\ \hline
\end{tabular}
\end{center}
\caption{ Numerical values for the  parameters,  the values are
given in units of $10^{-2}\,{\rm GeV^2}$ \cite{BFMS}.}
\end{table}
They are functions of eight independent parameters, $f_N$,
$\lambda_1$, $\lambda_2$, $V_1^d$, $A_1^u$, $f_d^1$, $f_d^2$ and
$f_u^1$,  the explicit expressions are presented in the appendix,
for detailed and systematic studies about this subject, one can
consult Ref.\cite{BFMS}.  Here we neglect the scale dependence and
take the following values for the eight independent parameters,
$f_N=(5.3\pm 0.5)\times 10^{-3} GeV^2$, $\lambda_1=-(2.7\pm
0.9)\times 10^{-2}GeV^2$, $\lambda_2=(5.1\pm 1.9)\times
10^{-2}GeV^2$, $V_1^d=0.23\pm 0.03$, $A_1^u=0.38\pm 0.15$,
$f_1^d=0.6\pm 0.2$, $f_2^d=0.15\pm 0.06$, $f_1^u=0.22\pm 0.15$.  In
estimating those parameters  with the QCD sum rules, only the first
few moments are taken into account, the values are not very
accurate.  One can map the uncertainties of the input parameters
into the uncertainties of the adjusted phenomenological parameters
\cite{Leinweber97,Lee97}, for example, the threshold parameter
$s_0$, the mass of the proton $M$, the scalar form-factor
$\sigma(t=-Q^2)$, etc,   with the Monte-Carlo method through
$\chi^2$ minimization, which may be the most realistic estimates of
the uncertainties as there are many input parameters which  can be
taken into account simultaneously, however, we are no expert in
Monte-Carlo simulation, the traditional uncertainties analysis is
chosen in this article.

We perform the operator product expansion in the light-cone with
large $Q^2$ and $P'^2$, the scalar form-factor $\sigma(t=-Q^2)$ make
sense at the range $Q^2>2GeV^2$, with the low momentum transfers,
the operator product expansion is questionable. In this article, we
devote to calculate the scalar form-factor at the range
$Q^2>2GeV^2$, which corresponding to the size about $0.1 fm$,
$(q-xP)^2 \rightarrow xM_B^2$ after the  Borel transformation, in
the region $M_B^2=(1.5-2.5)GeV^2$,  $\frac{1}{\sqrt{xM_B^2}} \leq
\frac{1}{\sqrt{0.5 \times 1.5} GeV} \sim 0.24 fm$,  retaining only
the three valence quark light-cone distribution amplitudes up to
twist-6 is reasonable.  The size of the proton is about the order of
$1fm$ which corresponding to the confinement scale
$\Lambda_{QCD}\approx 0.2GeV$, we only investigate short distance
physics inside the proton. With smaller momentum transfers, the
contributions from the soft (small virtual) gluons and quarks become
larger, and the multi-parton configurations become more and more
important, the quark and gluons degrees of freedom have to be
integrated  out,  we can work in the hadronic representation and
resort to the chiral perturbation theory to deal with the problems
\cite{BKM9396,Borasoy9679,OW99}. In numerical analysis, we observe
that  the scalar form-factor $\sigma(t=-Q^2)$ is sensitive to the
four parameters, $\lambda_1$, $f^d_1$, $f^d_2$ and $f^u_1$, which
are shown in Fig.3, Fig.4, Fig.5 and Fig.6, respectively.
\begin{figure}
 \centering
 \includegraphics[totalheight=7cm]{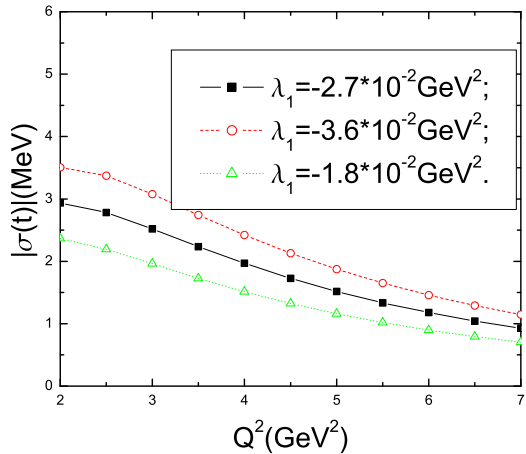}
 \caption{The   $\sigma(t)$ with the  $Q^2$ for the central values of the parameters except $\lambda_1$ . }
\end{figure}
\begin{figure}
 \centering
 \includegraphics[totalheight=7cm]{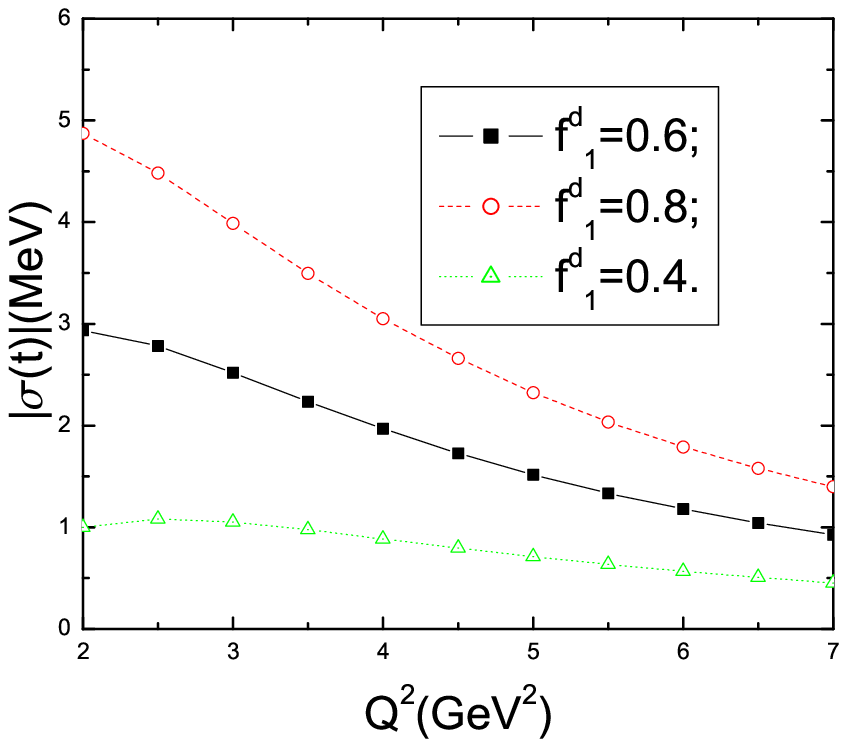}
 \caption{The   $\sigma(t)$ with the   $Q^2$ for the central values of the parameters except $f^d_1$. }
\end{figure}
\begin{figure}
 \centering
 \includegraphics[totalheight=7cm]{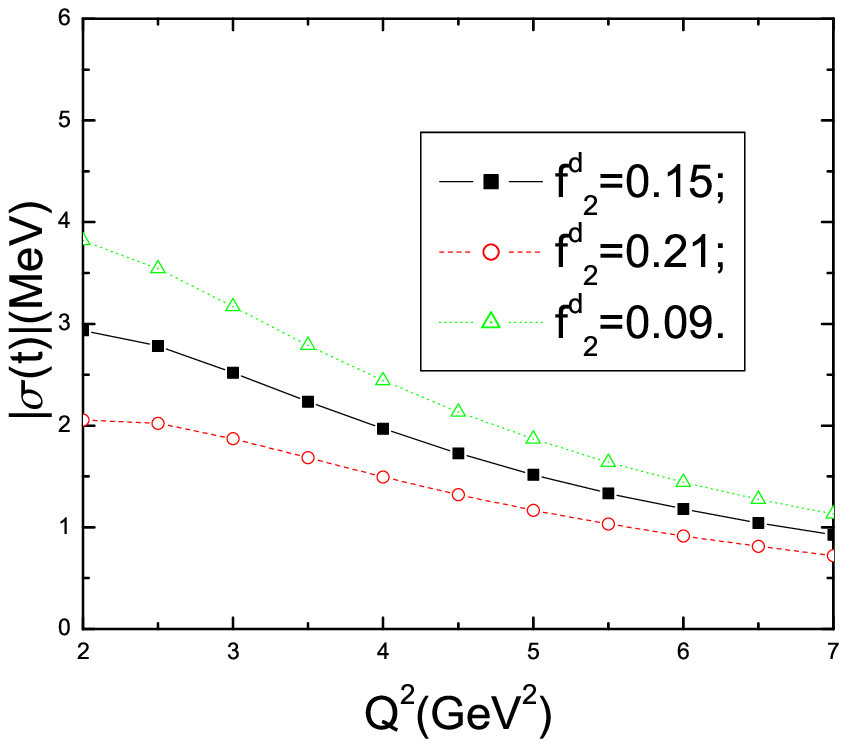}
 \caption{The   $\sigma(t)$ with the   $Q^2$ for the central values of the parameters except $f^d_2$. }
\end{figure}
\begin{figure}
 \centering
 \includegraphics[totalheight=7cm]{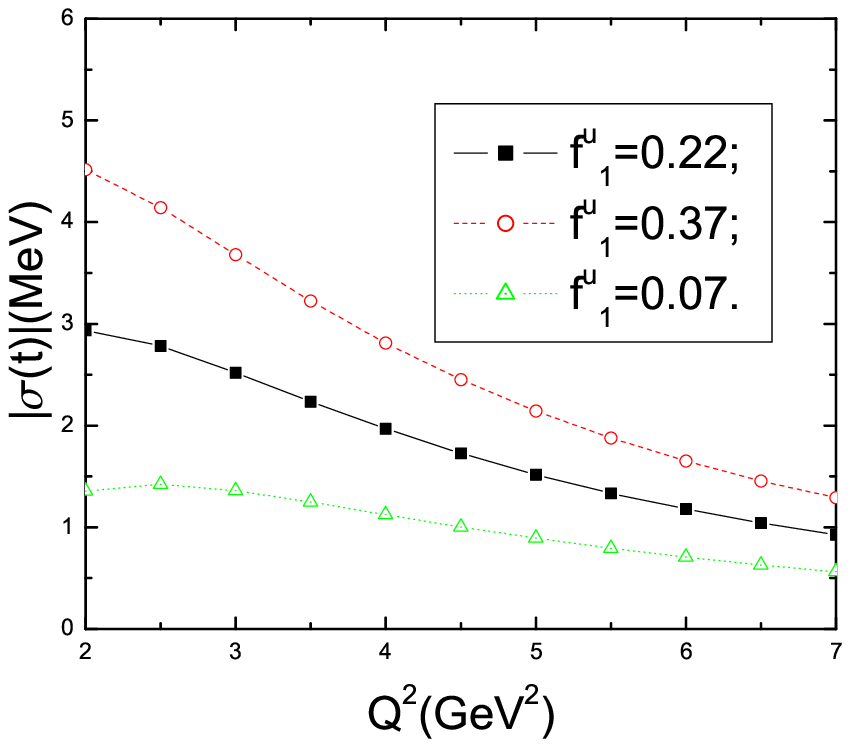}
 \caption{The   $\sigma(t)$ with the   $Q^2$ for the central values of the parameters except $f^u_1$. }
\end{figure}
Small variations of those parameters can lead to large changes for
the values, the large uncertainties can impair the predictive
 ability of the sum rules, and those parameters
$\lambda_1$, $f^d_1$, $f^d_2$ and $f^u_1$ should be refined to make
robust predications. The final numerical values for the scalar
form-factor $\sigma(t=-Q^2)$ at the intermediate and large
space-like momentum regions $Q^2>2GeV^2$ are plotted in the Fig.7,
from the figure, we can see that the values of the scalar
form-factor $\sigma(t=-Q^2)$ are compatible with the calculations of
lattice QCD \cite{DLL96} and chiral quark models \cite{ChiralForm}.
If we take the values from the recent analysis of the $\pi N$
scattering data, $\sigma(0)\approx (\Sigma_{\pi N}=79\pm7MeV)$, as
input, the results from the lattice calculation give
$\sigma(-2GeV^2)<\sigma(0)\times 0.1\approx 7.9MeV$, our results
$|\sigma(-2GeV^2)|\approx (2.9\pm 2.7) MeV $ are smaller, however,
reasonable. The $\alpha_s$ corrections to the scalar form-factor of
the proton may be significant, quantitative conclusion can be
reached after the solid calculations, the calculations are tedious
though not impossible, and beyond the present work. In the case of
the $\pi$ meson, the $\alpha_s$ corrections of the twist-2
light-cone distribution amplitude reproduce the $\frac{1}{Q^2}$
behavior for the electro-magnetic form-factor with large $Q^2$,
which corresponding to the hard re-scattering mechanism
\cite{BKM00}. The contributions from the four-particle (and
five-particle) nucleon distribution amplitudes with a gluon or
quark-antiquark pair in addition to the three valence quarks are
usually not expected to play any significant roles \cite{DFJK} and
neglected here.  The consistent and complete LCSR analysis should
include the contributions from the perturbative $\alpha_s$
corrections, the distribution amplitudes with additional valence
gluons and quark-antiquark pairs, and improve the parameters which
enter in the LCSRs.
\begin{figure}
 \centering
 \includegraphics[totalheight=7cm]{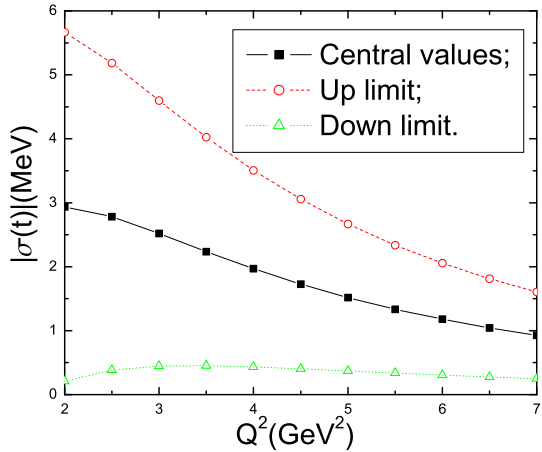}
 \caption{The   $\sigma(t)$ with the  $Q^2$. }
\end{figure}
\section{Conclusion }

In this work, we  calculate the
 scalar form-factor $\sigma(t=-Q^2)$ of the proton  in the framework of the LCSR approach up
 to twist-6 three valence quark light-cone distribution amplitudes
  and observe the scalar form-factor $\sigma(t=-Q^2)$ with  intermediate and
large momentum transfers, $Q^2> 2 GeV^2$, has significant
contributions from the end-point (or soft) terms, it is consistent
with the growing consensus that the onset of the perturbative QCD
region in exclusive processes is postponed to very large energy
scales.  In numerical analysis, we observe that the scalar
form-factor $\sigma(t=-Q^2)$ is sensitive to the four parameters,
$\lambda_1$, $f^d_1$, $f^d_2$ and $f^u_1$, small variations of those
parameters can lead to large changes for the values. The large
uncertainties can impair the predictive ability of the sum rules,
the parameters $\lambda_1$, $f^d_1$, $f^d_2$ and $f^u_1$ should be
refined to make robust predications. The numerical values for the
$\sigma(t=-Q^2)$ are
 compatible with the calculations from the chiral quark model and lattice QCD.
  The consistent and complete
LCSR analysis should include the contributions from the perturbative 
$\alpha_s$ corrections, the distribution amplitudes with additional
valence gluons and quark-antiquark pairs, and improve the parameters
which enter in the LCSRs.

\section*{Acknowledgment}
This  work is supported by National Natural Science Foundation,
Grant Number 10405009,  and Key Program Foundation of NCEPU. The
authors are indebted to Dr. J.He (IHEP), Dr. X.B.Huang (PKU) and Dr.
L.Li (GSCAS) for numerous help, without them, the work would not be
finished.
\appendix
\section*{Appendix}
\begin{eqnarray}
\phi_3^0 = \phi_6^0 = f_N \,,\hspace{0.3cm} &\qquad& \phi_4^0 =
\phi_5^0 = \frac{1}{2} \left(\lambda_1 + f_N\right) \,,
\nonumber \\
\xi_4^0 = \xi_5^0 = \frac{1}{6} \lambda_2\,, &\qquad& \psi_4^0  =
\psi_5^0 = \frac{1}{2}\left(f_N - \lambda_1 \right)  \,. \nonumber
\end{eqnarray}
\begin{eqnarray}
\tilde\phi_3^- &=& \frac{21}{2} A_1^u,\nonumber\\
\tilde\phi_3^+ &=& \frac{7}{2} (1 - 3 V_1^d),\nonumber\\
\phi_4^- &=& \frac{5}{4} \left(\lambda_1(1- 2 f_1^d -4 f_1^u) + f_N(
2 A_1^u - 1)\right) \,,
\nonumber \\
\phi_4^+ &=& \frac{1}{4} \left( \lambda_1(3- 10 f_1^d) - f_N( 10
V_1^d - 3)\right)\,,
\nonumber \\
\psi_4^- &=& - \frac{5}{4} \left(\lambda_1(2- 7 f_1^d + f_1^u) +
f_N(A_1^u + 3 V_1^d - 2)\right) \,,
\nonumber \\
\psi_4^+ &=& - \frac{1}{4} \left(\lambda_1 (- 2 + 5 f_1^d + 5 f_1^u)
+ f_N( 2 + 5 A_1^u - 5 V_1^d)\right)\,,
\nonumber \\
\xi_4^- &=& \frac{5}{16} \lambda_2(4- 15 f_2^d)\,,
\nonumber \\
\xi_4^+ &=& \frac{1}{16} \lambda_2 (4- 15 f_2^d)\,,\nonumber\\
\phi_5^- &=& \frac{5}{3} \left(\lambda_1(f_1^d - f_1^u) + f_N( 2
A_1^u - 1)\right) \,,
\nonumber \\
\phi_5^+ &=& - \frac{5}{6} \left(\lambda_1 (4 f_1^d - 1) + f_N( 3 +
4 V_1^d)\right)\,,
\nonumber \\
\psi_5^- &=& \frac{5}{3} \left(\lambda_1 (f_1^d - f_1^u) + f_N( 2 -
A_1^u - 3 V_1^d)\right)\,,
\nonumber \\
\psi_5^+ &=& -\frac{5}{6} \left(\lambda_1 (- 1 + 2 f_1^d +2 f_1^u) +
f_N( 5 + 2 A_1^u -2 V_1^d)\right)\,,
\nonumber \\
\xi_5^- &=& - \frac{5}{4} \lambda_2 f_2^d\,,
\nonumber \\
\xi_5^+ &=&  \frac{5}{36} \lambda_2 (2 - 9 f_2^d)\,,
\nonumber \\
\phi_6^- &=& \phantom{-}\frac{1}{2} \left(\lambda_1 (1- 4 f_1^d - 2
f_1^u) + f_N(1 +  4 A_1^u )\right) \,,
\nonumber \\
\phi_6^+ &=& - \frac{1}{2}\left(\lambda_1  (1 - 2 f_1^d) + f_N ( 4
V_1^d - 1)\right)\,. \nonumber
\end{eqnarray}

\end{document}